\documentclass[letter]{spie}  

\usepackage[dvips]{graphicx}
\usepackage{overcite}
\usepackage{bm}

\title{Coupling Efficiencies in Single Photon On-Demand Sources}

\author{Stefania Castelletto\supit{a},I.P. Degiovanni\supit{a}, Michael
Ware\supit{b} Alan Migdall\supit{b} \\
\footnotesize{\supit{\tiny{a}}Istituto Elettrotecnico Nazionale G.
Ferraris,
Strada delle Cacce 91-10135 Torino (Italy), \\
\supit{\tiny{b}}Optical Technology Division, National Institute of
Standards and Technology,Gaithersburg, Maryland 20899-8441 }}

\authorinfo{
Send correspondence to Alan Migdall: Email:amigdall@nist.gov,
Telephone: 001 301 975 2331}

\begin{document}
   \maketitle


\begin{abstract}

Many quantum computation and communication schemes require, or
would significantly benefit from, true sources of single photon
on-demand (SPOD). Unfortunately, such sources do not exist. It is
becoming increasingly clear that coupling photons out of a SPOD
source will be a limiting factor in many SPOD implementations. In
particular, coupling these source outputs into optical fibers
(usually single mode fibers) is often the preferred method for
handling this light.  We investigate the practical limits to this
coupling as relates to parametric downconversion, an important
starting point for many SPOD schemes. We also explored whether it
is possible to optimize the engineering of the downconversion
sources to improve on this coupling. We present our latest results
in this area.

\end{abstract}



\section{\label{sec:level1} Introduction}

The advent of photon-based quantum cryptography, communication,
and computation
schemes\cite{BEB84,BEB85,BEB89,BBB91,EKE91,BEN92,BBM92,ERT92,TBZ00,KLM01}
has increased the need for light sources that produce individual
photons\cite{BLM00}. An ideal single-photon source would produce
completely characterized single photons on demand. Since all of
the currently available sources fall far short of this ideal, many
new schemes have been proposed for creating single photon on
demand sources (SPOD). Many of these schemes\cite{PJF02,MBC02}
rely on optical parametric downconversion (PDC), because it
produces photons two at a time, allowing one photon to herald the
existence of the other.  In a previous work we proposed one such
scheme for a multiplexed single photon on-demand source that
increases the probability of single photon emission while
suppressing the probability of multi-photon
generation\cite{MBC02}. Most PDC based schemes (including ours)
require that the PDC output be collected into a single spatial
mode defined by an optical fiber.  In order for these PDC schemes
to produce a reliable SPOD source, it is essential that the
optical collection system efficiently gathers and detects the
herald photon and sends its twin to the output path of the system.

Various theoretical models have been developed to predict how the
collection efficiency of PDC light in a ``two-photon single mode''
can be improved \cite{MRP98a,KOW01,BGS03}, and in many cases these
models have been experimentally proven to give a better coupling
efficiency. In particular, it has been shown that the position and
size of the pump beam focus affect the shape of the PDC
output\cite{MRP98a} and hence the coupling efficiency of the PDC
into a given spatial mode.  In addition, a more detailed
work\cite{BGS03} recently showed how crystal length and walk-off
also affect the coupling efficiency for a pulsed broadband pump.
(However, in that recent work the attempts to increase the single
mode fiber coupling efficiency were centered on having a thin
crystal and a tightly focused pump beam, which is achieved at a
cost of overall source brightness.) The work done to date on
coupling efficiency gives some guidance for increasing coupling
efficiency.  However, there is still not a practical ``recipe'' to
quantify and maximize the coupling of PDC light into single mode
fibers.

In this paper we present a method for analytically calculating the
coupling efficiency of PDC light into single mode fiber.
Specifically, we calculate the spatial and angular overlap between
type~I PDC output and the two single modes defined by optical
fibers. The collection efficiency is evaluated in terms of pump
waist, collecting mode, crystal length, and the imaging conditions
of the optical collection system. To ease the computational
difficulties associated with this problem, we make the
approximation of near-perfect phase matching.  We quantify the
limitations introduced by this approximation and then show by
example how we can obtain better coupling efficiency by properly
shaping the pump beam and collection modes for a given crystal
length. Although our approximations limit the length of crystals
that can be analyzed using this method, this formalism does allow
us to study crystals with lengths beyond what is considered the
thin crystal limit. As a next step in this research we plan to use
a more exact model with a numerical evaluation that allows us to
model the efficiency over a wider range of parameter values
(expecially longer crystal cases).

\section{\label{sec:level1} Theory}

To determine the coupling efficiency between the parametric
downconversion output and two single collection modes (defined in
our setup by single-mode optical fibers) we must calculate the
spatial overlap of the two-photon field of the PDC source with the
modes selected by the fibers.  The first step in this procedure is
to calculate the two-photon field \cite{RUB96} given by
\begin{equation}
    A_{12} = \left< 0 \right| E_1^{(+)} E_2^{(+)} \left| \psi
    \right>, \label{eqn:biphoton}
\end{equation}
where $\left| 0 \right>$ is the vacuum state and $E_1^{(+)}$ and
$E_2^{(+)}$ are the positive-frequency portions of the electric
field evaluated at positions $\mathbf{r}_1$ and $\mathbf{r}_2$.
The two-photon state at the output surface of a PDC crystal
oriented with its face perpendicular to the $z$-axis is given by
\begin{equation}\label{1}
    \left| \psi \right> =
    \int\textrm{d}^{3}k_\mathrm{s}\int\textrm{d}^{3}k_\mathrm{i}\int\textrm{d}^{2}
    \kappa_\mathrm{p} \int_{S} \textrm{d}^{2}\rho \int_{0}^{L}\textrm{d}z \;
    \widetilde{E}_\mathrm{p}(\bm{\kappa}_\mathrm{p})
    e^{i(\Delta k_z z + \Delta \bm{\kappa} \cdot\bm{\rho})}
    |1_{\mathbf{k}_\mathrm{s}}
    \rangle |1_{\mathbf{k}_\mathrm{i}} \rangle ,
\end{equation}
where $S$ is the cross sectional area of the crystal illuminated
by the pump, $\bm{\rho}$ is the transverse distance from the
$z$-axis, and $L$ is the length of the crystal. We designate the
component of a wavevector that is parallel to the crystal face by
$\bm{\kappa}$, so that $\mathbf{k}_j = {k_z}_j \hat{z} +
\bm{\kappa}_j$. The subscripts $j=s$, $i$, $p$ indicate the
signal, idler, and pump. The quantities $\Delta k_z =
{k_\mathrm{p}}_{z} - {k_\mathrm{s}}_{z} - {k_\mathrm{i}}_{z}$ and
$\Delta \bm{\kappa} = \bm{\kappa}_\mathrm{p} -
\bm{\kappa}_\mathrm{i} - \bm{\kappa}_\mathrm{s} $ are referred to
as the longitudinal and transverse wave-vector mismatch,
respectively. The pump beam's angular spectrum defines its
transverse distribution via the Fourier transform
\begin{equation}\label{2}
    E_\mathrm{p} (\bm{\rho})= \frac{1}{2 \pi} \int\textrm{d}^2 \kappa_\mathrm{p} ~
    \widetilde{E}_\mathrm{p}(\bm{\kappa}_\mathrm{p} )
    e^{ i \bm{\kappa}_\mathrm{p} \cdot \bm{\rho} } .
\end{equation}
This analysis assumes that the pump propagates with negligible
diffraction effect inside the crystal, so that $E_\mathrm{p}
(\bm{\rho})$ is independent of $z$.

The electric fields in Eq.~(\ref{eqn:biphoton}) are evaluated
outside the crystal, while the wavevectors in Eq.~(\ref{1}) are
evaluated inside the crystal.  To evaluate
Eq.~(\ref{eqn:biphoton}) it is convenient to write $\Delta k_z$ in
terms of the angular frequency, $\omega$, and the index of
refraction, $n(\omega)$, since the $z$ components of the
wavevectors are discontinuous at the crystal surface. Thus, the
$z$~component of a wavevector is
\begin{equation}
  k_z = \sqrt{\left(\frac{ n(\omega) \omega}{c}\right)^2 -
  \kappa^2} , \label{eqn:kz}
\end{equation}
where $c$ is the speed of light.  We assume that the pump beam has
a narrow angular spectrum (transverse wavevector distribution) and
that the signal and idler are observed only at points close to the
$z$ axis.  In addition, we assume that only a narrow range of
frequencies are collected because narrow bandpass filters are
placed the signal and idler optical paths. The central frequencies
defined be these filters are specified by $\Omega_\mathrm{s}$ and
$\Omega_\mathrm{i}$. (In our notation we denote central
frequencies chosen as system parameters by capital $\Omega$ and
frequency variables by lower case $\omega$.) We then expand
$\Delta k_z$ around these central frequencies to obtain
\cite{RUB96}
\begin{equation}\label{8}
    \Delta k_z \approx \Delta k_z^o + D \omega-D' \omega^{2} +
    \frac{c}{2~n_\mathrm{s}(\Omega_\mathrm{s})\Omega_\mathrm{s} \cos\theta_1 } \kappa_\mathrm{s}^2
    + \frac{c}{2~n_\mathrm{i}(\Omega_\mathrm{i})\Omega_\mathrm{i} \cos\theta_2 } \kappa_\mathrm{i}^2
    - \frac{c}{2~n_\mathrm{p}(\theta_\mathrm{p}, \Omega_\mathrm{p})\Omega_\mathrm{p}}
    \kappa_\mathrm{p}^2,
\end{equation}
where we have
\begin{eqnarray}
    \Delta k_z^o  & = & \frac{n_\mathrm{p}(\theta_\mathrm{p}, \Omega_\mathrm{p}) \Omega_\mathrm{p}}{c} -
    \frac{n_\mathrm{s}(\Omega_\mathrm{s})\Omega_\mathrm{s}\cos\theta_1}{c  } -
    \frac{n_\mathrm{i}(\Omega_\mathrm{i})\Omega_\mathrm{i}\cos\theta_2}{c  } \\
    \nonumber
    D & = &
    \left. \frac{\textrm{d} n_\mathrm{i}(\omega_\mathrm{i})\omega_\mathrm{i}/c}{\textrm{d}\omega_\mathrm{i}}
    \right|_{\Omega_\mathrm{i}}
    -\left. \frac{\textrm{d} n_\mathrm{s}(\omega_\mathrm{s})\omega_\mathrm{s}/c}{\textrm{d}\omega_\mathrm{s}}
    \right|_{\Omega_\mathrm{s}}
    \\ \nonumber
    D'& = & \left. \frac{\textrm{d}^2 n_\mathrm{i}(\omega_\mathrm{i})\omega_\mathrm{i}/c} {\textrm{d} \omega_\mathrm{i}^2}
    \right|_{\Omega_\mathrm{i}} + \left. \frac{\textrm{d}^2 n_\mathrm{s}(\omega_\mathrm{s})\omega_\mathrm{s}/c}{\textrm{d}
    \omega_\mathrm{s}^2} \right|_{\Omega_\mathrm{s}} . \\ \nonumber
\end{eqnarray}
In obtaining these expressions we have assumed type-I
phasematching with perfect frequency matching, i.e.
$\Omega_\mathrm{p}=\Omega_\mathrm{i}+\Omega_\mathrm{s}$  where
$\Omega_\mathrm{p}$ is the pump angular frequency. The angle
between the pump propagation direction and the crystal optic axis
is specified by $\theta_\mathrm{p}$, and the PDC photons emission
angles are given by $\theta_1 $ and $\theta_2 $. We have developed
the angular frequency as $\omega_\mathrm{s}= \Omega_\mathrm{s} +
\omega $ and $\omega_\mathrm{i}= \Omega_\mathrm{i} + \omega' $,
with $\omega'=-\omega$.  Note that in the degenerate case we have
$D=0$.

We assume the PDC light is collected into two single-mode fibers
using identical lenses (with focal lengths $f$) both placed at a
distances $d$ from the fiber tips and $M d$ from the crystal face,
where $M$ is the magnification. (We assume that the fibers define
single Gaussian modes. See Fig.~1 for a schematic of this setup.)
In this situation we can write the biphoton field as
\begin{eqnarray}\label{4}
    A_{12}(\bm{\rho}_1,
    \bm{\rho}_2)&\propto&
    \int\textrm{d}\omega \int\textrm{d}^{2} \kappa_\mathrm{s}
    \int\textrm{d}^{2} \kappa_\mathrm{i}
    \int\textrm{d}^{2} \kappa_\mathrm{p}
    \widetilde{E}_\mathrm{p} (\bm{\kappa}_\mathrm{p})
    \int_{S}\textrm{d}^{2}\rho \int_{0}^{L}\textrm{d}z \\ & &
    \nonumber \times  ~
    e^{i (\Delta k_z z + \Delta \bm{\kappa} \cdot \bm{\rho})}
    H_\mathrm{i}(\bm{\kappa}_\mathrm{i}, \bm{\rho}_{1})
    H_\mathrm{s}(\bm{\kappa}_\mathrm{s}, \bm{\rho_{2}})
    \tau_{1}(\omega) \tau_{2}(\omega),
\end{eqnarray}
where $H_\mathrm{i}(\bm{\kappa}_\mathrm{i}, \bm{\rho}_{1}) $ and
$H_\mathrm{s}(\bm{\kappa}_\mathrm{s}, \bm{\rho}_2) $ are the
transfer functions of the lenses in the optical path of the idler
and signal and $\bm{\rho}_1$ and $\bm{\rho}_2$ are the transverse
coordinates measured from the $\hat{z}_1$ and $\hat{z}_2$ axes,
respectively (see Fig.~1), at the imaging plane of the lenses. The
functions $\tau_1(\omega)$ and $\tau_2(\omega)$ define the
spectral transmittances of interference filters placed in the two
collection paths. To simplify the form of the transfer functions,
we use the paraxial approximation and consider an ideal lens with
infinite aperture.  Under these assumptions the transfer functions
are given by
\begin{eqnarray}
    H_\mathrm{i}( \bm{\kappa}_\mathrm{i} , \bm{\rho}_1 ) & \propto &
    e^{i M \bm{\kappa}_\mathrm{i} \cdot \bm{\rho}_1}
    e^{i c \epsilon d^2 \kappa_\mathrm{i}^2 / (2 \Omega_\mathrm{i})  } \\ \nonumber
    H_\mathrm{s}( \bm{\kappa}_\mathrm{s} , \bm{\rho}_2) & \propto &
    e^{i M \bm{\kappa}_\mathrm{s} \cdot \bm{\rho}_2}
    e^{i c \epsilon d^2 \kappa_\mathrm{s}^2 / ( 2 \Omega_\mathrm{s} )  }
\end{eqnarray}
where $\epsilon = 1/d +1/(M d) - 1/f$ provides for non-perfect
imaging caused by precision limits in the setting of the
lens-fiber-crystal distances.

\begin{figure}[tbp]
\par
\begin{center}
\includegraphics{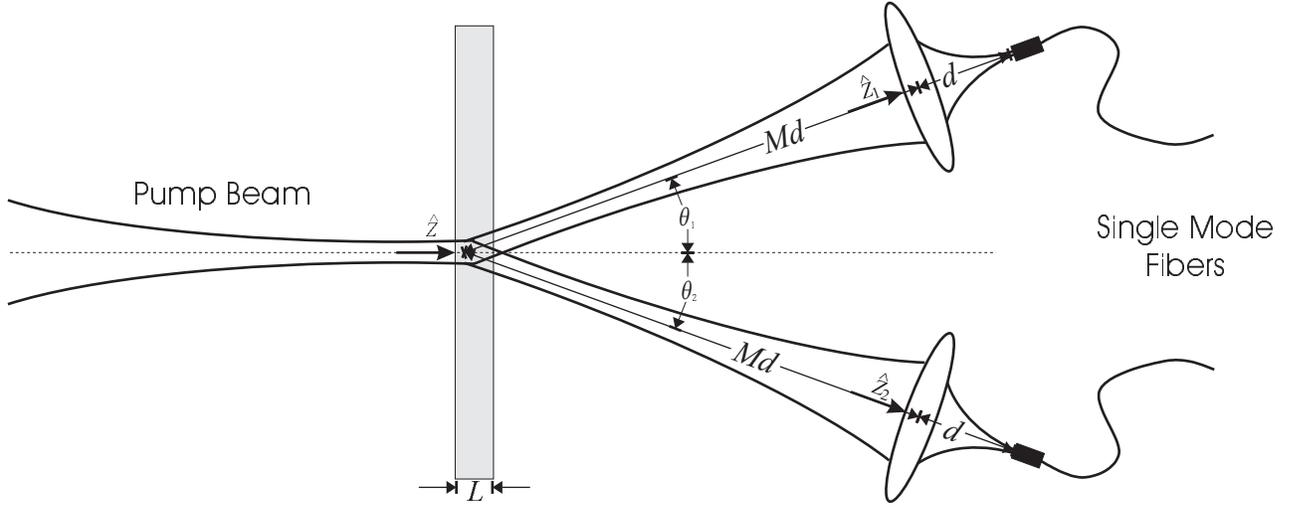}
\end{center}
\caption{Schematic of collection setup.}
\end{figure}

The pump field transverse distribution is defined to be Gaussian
with a waist of $\mathcal{W}_\mathrm{p}$, and we assume that the
transverse crystal size is large relative to the pump beam, so we
can take the cross section $S$ to be infinite. We also assume
perfect transverse phase matching, i.e. $\bm{\kappa}_\mathrm{p} =
\bm{\kappa}_\mathrm{s} + \bm{\kappa}_\mathrm{i}$. These
assumptions allow the $\bm{\kappa}_\mathrm{p}$ and $\bm{\rho}$
integrals in Eq.~(\ref{4}) to be trivially performed
\begin{equation} \label{eqn:pumpint}
    \int_{S}\textrm{d}^{2}\rho \left(
    \int\textrm{d}^{2} \kappa_\mathrm{p} ~
    \widetilde{E}_\mathrm{p} (\bm{\kappa}_\mathrm{p})
    e^{i (\bm{\kappa}_\mathrm{p} \cdot \bm{\rho})} \right)
    e^{-i (\bm{\kappa}_\mathrm{s} + \bm{\kappa}_\mathrm{i}) \cdot \bm{\rho}}
    ~ \propto ~ \widetilde{E}_\mathrm{p} ( \bm{\kappa}_\mathrm{i} + \bm{\kappa}_\mathrm{s}
    ).
\end{equation}
We next assume that the longitudinal mismatch is small ($\Delta
k_z L \ll 1$) and that we have near-perfect imaging ($c \epsilon
d^{2}\kappa_{s,i}^2 /(2 \Omega_{s,i}) \ll 1$). In this case, we
can expand the exponentials in the Eq.~(\ref{4}) and use the
result in Eq.~(\ref{eqn:pumpint}) to obtain
\begin{eqnarray}\label{7}
    A_{12}(\bm{\rho}_1, \bm{\rho}_2)
    &\propto&
    ~ \int\textrm{d}\omega \int\textrm{d}^{2} \kappa_\mathrm{s}
    \int\textrm{d}^{2} \kappa_\mathrm{i}
    \widetilde{E}_\mathrm{p} ( \bm{\kappa}_\mathrm{i} + \bm{\kappa}_\mathrm{s} )
    e^{ i M ( \bm{\kappa}_\mathrm{s} \cdot \bm{\rho}_2 +
    \bm{\kappa}_\mathrm{i} \cdot \bm{\rho}_1)} \tau_1(\omega) \tau_2(\omega)
    \\ & & \nonumber \times ~
    \left( 1+ \frac{i}{2} c  \epsilon d^2 \left(
    \frac{\kappa_\mathrm{s}^2}{\Omega_\mathrm{s}}
    + \frac{\kappa_\mathrm{i}^2}{\Omega_\mathrm{i}} \right) + \cdots \right)
    \left( 1 + \frac{i}{2}  \Delta k_z L -\frac{1}{6} \left( \Delta k_z L\right)^2+ \cdots \right)
\end{eqnarray}
The spectral filters defined by $\tau_1$ and $\tau_2$ are centered
around the wavelengths $\Omega_\mathrm{s}$ and $\Omega_\mathrm{i}$
are both assumed to have a rectangular bandwidth $B$. We limit the
expansions to first order and assume degenerate type-I
phase-matching (i.e. $D=0$). The integrals yield
\begin{equation}\label{11}
    A_{12}(\bm{\rho}_1, \bm{\rho}_2 ) \propto
     B A^o_{12} (\bm{\rho}_1 ,\bm{\rho}_2)
    \left[ 1 +\frac{i L}{2} \left( \alpha+ \beta ~ a_{1} + \gamma ~ a_{2} -
    ~ \nu ~ a_{3} \right) \right]
\end{equation}
where the zero order solution for the biphoton field is
\begin{equation}\label{10}
    A^o_{12}( \bm{\rho}_1, \bm{\rho}_2 ) \propto
    \frac{1}{M^2}
    E_\mathrm{p} \left( \frac{(\bm{\rho}_1 + \bm{\rho}_2 )M}{2} \right)
    \delta( \bm{\rho}_1 - \bm{\rho}_2 )
    \int\textrm{d}\omega \; \tau_{1}(\omega) \tau_{2}(\omega).
\end{equation}
and the coefficients are
\begin{eqnarray}
    \alpha & = & \Delta k_z^o -D' B^2/12 \\ \nonumber
    \beta & = &
    \frac{c}{2~ n_\mathrm{i} (\Omega_\mathrm{i} ) \Omega_\mathrm{i} \cos\theta_2} +
    \frac{c \epsilon d^2}{ \Omega_\mathrm{i} L} -
    \frac{c}{2~n_\mathrm{p}(\Omega_\mathrm{p},\theta_\mathrm{p})\Omega_\mathrm{p} } \\ \nonumber
    \gamma & = &
    \frac{c}{2~n_\mathrm{s} (\Omega_\mathrm{s}) \Omega_\mathrm{s} \cos\theta_1} +
    \frac{c \epsilon~d^2}{\Omega_\mathrm{s} L} -
    \frac{c}{2~ n_\mathrm{p}(\Omega_\mathrm{p},\theta_\mathrm{p}) \Omega_\mathrm{p} } \\ \nonumber
    \nu & = &
    \frac{c}{2~n_\mathrm{p}(\Omega_\mathrm{p},\theta_\mathrm{p}) \Omega_\mathrm{p} } \\ \nonumber
    a_1 & = &
    4 \frac{ (1-M^2 \rho_1^2 / \mathcal{W}_\mathrm{p}^2)}{ \mathcal{W}_\mathrm{p}^2 } \\ \nonumber
    a_2 & = &
    4 \frac{ (1-M^2 \rho_2^2 / \mathcal{W}_\mathrm{p}^2) }{ \mathcal{W}_\mathrm{p}^2} \\ \nonumber
    a_3 & = &
    2 (\frac{2 y_2 \delta' (y_1 -y_2)}{ \mathcal{W}_\mathrm{p}^2} +
    \frac{ \delta'' ( y_1 - y_2 ) }{M^2}),
\end{eqnarray}

With Eq.~(\ref{11}) in hand we are prepared to calculate the
coupling of the biphoton field into the single mode fibers in our
setup. To accomplish this we write $A_{12}$ as a coherent
superposition of guided modes in the fiber as suggested in
Ref.~\citen{BGS03}
\begin{equation}
   A_{12}(\bm{\rho}_1,\bm{\rho}_2) =
   \sum_{l' m',l m}{ A_{12}}^{l' m',l m}
   \varphi^{*}_{l'm'}(\bm{\rho}_1)
   \varphi^{*}_{lm}(\bm{\rho}_2)
\end{equation}
The collection efficiency can then be calculated by
\begin{equation}\label{12}
   \eta_{12} = \frac{C_{12}}{ \sqrt{C_1 C_2} },
\end{equation}
where $C_{12}$, $C_1$, and $C_2$ measure the square of the overlap
between the biphoton field and the collection modes, and can be
used to calculate singles and coincidence counting rates
\begin{eqnarray}
    C_{12} & = & \left| \int \textrm{d}^2 \rho_1 \int \textrm{d}^2 \rho_2
     ~  A_{12} ( \bm{\rho}_1, \bm{\rho}_2)
    \varphi^{*}_{l'm'} (\bm{\rho}_1)
    \varphi^{*}_{lm} ( \bm{\rho}_2) \right|^2 \\ \nonumber
    C_1 & = & \int \textrm{d}^2 \rho_2 \left| \int\textrm{d}^2\rho_1
     ~  A_{12} ( \bm{\rho}_1 , \bm{\rho}_2 )
    \varphi^{*}_{l'm'} (\bm{\rho}_1) \right|^2 \\ \nonumber
    C_2 & = & \int \textrm{d}^2\rho_1 \left| \int\textrm{d}^2\rho_2
     ~  A_{12}( \bm{\rho}_1 , \bm{\rho}_2 )
    \varphi^{*}_{l m} (\bm{\rho}_2) \right|^2 .
\end{eqnarray}
We assume the guided mode is a gaussian defined by
\begin{equation}
    \varphi^{*}_{10} (\bm{\rho}) = \sqrt{\frac{2}{\pi}}
    \frac{1}{\mathcal{W}_f} \exp \left[ - \frac{\rho^2}{\mathcal{W}_f^2} \right]
\end{equation}
where $\mathcal{W}_f$ is the width of the collection fiber.  The
waist of the collected mode imaged back to the crystal is given by
$\mathcal{W}$, where $\mathcal{W}=M \mathcal{W}_f$. The collection
efficiency is then obtained by using Eq.(\ref{11}) and (\ref{12})
\begin{equation}
    \label{eqn:efficiency}
    \eta_{12} \simeq \eta_{12}^{o}
    \frac{1+\frac{1}{4} L^2 \chi_1^2}{1+ \frac{1}{4} L^2 \chi_2}
\end{equation}
where the collection efficiency at the zero order of the
perturbation theory (very thin crystals) is
\begin{equation}\label{16}
    \eta_{12}^{o} = 4 \frac{ \mathcal{W}^2 / \mathcal{W}_\mathrm{p}^2 + 1}{(\mathcal{W}^2/\mathcal{W}_\mathrm{p}^2 + 2)^2}
\end{equation}
and the corrections are
\begin{eqnarray}
    \chi_{1} & = & \frac{
    \int \textrm{d}^2 \rho_2 \int \textrm{d}^2 \rho_1  ~
    A_{12}^{o}(\bm{\rho}_1 , \bm{\rho}_2)
    \varphi^{*}_{10}(\bm{\rho}_1) \varphi^{*}_{10} (\bm{\rho}_2)
    ( \alpha + \beta a_1 + \gamma a_2 - \nu  a_3)}
    { \int \textrm{d}^2\rho_2 \int \textrm{d}^2 \rho_1
    A_{12}^{o}(\bm{\rho}_1 , \bm{\rho}_2)
    \varphi^{*}_{10}(\bm{\rho}_1) \varphi^{*}_{10} (\bm{\rho}_2)}
    \\ \nonumber
    \chi_{2} & = & \frac{\int \textrm{d}^2 \rho_2 \left|
    \int \textrm{d}^2 \rho_1  ~
    A_{12}^{o} (\bm{\rho}_1 , \bm{\rho}_2)
    \varphi^{*}_{10}( \bm{\rho}_1 )
    (\alpha + \beta a_1 + \gamma a_2 - \nu a_3) \right|^2}
    {\int \textrm{d}^2 \rho_2 \left| \int \textrm{d}^2 \rho_1
    A_{12}^{o}( \bm{\rho}_1 , \bm{\rho}_2 )
    \varphi^{*}_{10}( \bm{\rho}_1) \right|^2 }
\end{eqnarray}
We can perform these integrals using the expressions developed
above to obtain
\begin{eqnarray}
    \chi_1 & = & \alpha + \frac{2 \nu}{\mathcal{W}^2} +
    \frac{8 \beta - 2 \nu + 8 \gamma}{ \mathcal{W}^2 + 2~ \mathcal{W}_\mathrm{p}^2}
    \\ \nonumber
    \chi_{2} & = & \alpha^2 + \frac{12 \nu^2}{\mathcal{W}^4} +
    \frac{16 (\beta+ \gamma) \nu }{ \mathcal{W}^2 \mathcal{W}_\mathrm{p}^2} +
    ( \beta + \gamma)^2 (\frac{1}{\mathcal{W}_\mathrm{p}^4} +
    \frac{1}{( \mathcal{W}^2 + \mathcal{W}_\mathrm{p}^2)^2}) + 4 \alpha(\frac{\nu}{\mathcal{W}^2} +
    (\beta + \gamma)(\frac{1}{\mathcal{W}_\mathrm{p}^2} + \frac{1}{\mathcal{W}^2 + \mathcal{W}_\mathrm{p}^2})) .
\end{eqnarray}

\section{\label{sec:level1} Examples}

To illustrate the use of Eq.~(\ref{eqn:efficiency}) we consider a
Lithium Iodate crystal with $L=2~\mathrm{mm}$.  The crystal is
pumped with 458~nm light with $\theta_\mathrm{p}=36^{\circ}$ and
the collection modes are arranged to collect degenerate PDC
emission at $\theta_1 = \theta_2 = 1.5^\circ$ with a bandwidth of
$B=2 \times 10^{13}~\mathrm{s}^{-1}$ (about 10 nm). The
magnification is fixed at $M=20$ and we assume that the collection
mode waist occurs at the crystal with perfect imaging
($\epsilon=0$). Figure~2a plots the collection efficiency versus
the pump waist for different collection mode waists. Figure~2b
plots the term $\Delta k_z L$, and Figs.~2c and 2d plot the
correction factors $\chi_1^2 L^2/4$ and $\chi_2 L^2/4$,
respectively.  Notice that the correction factors become
significant for small pump waists (i.e. broad angular spectra of
the pump).

\begin{figure}[tbp]
\par
\begin{center}
\includegraphics{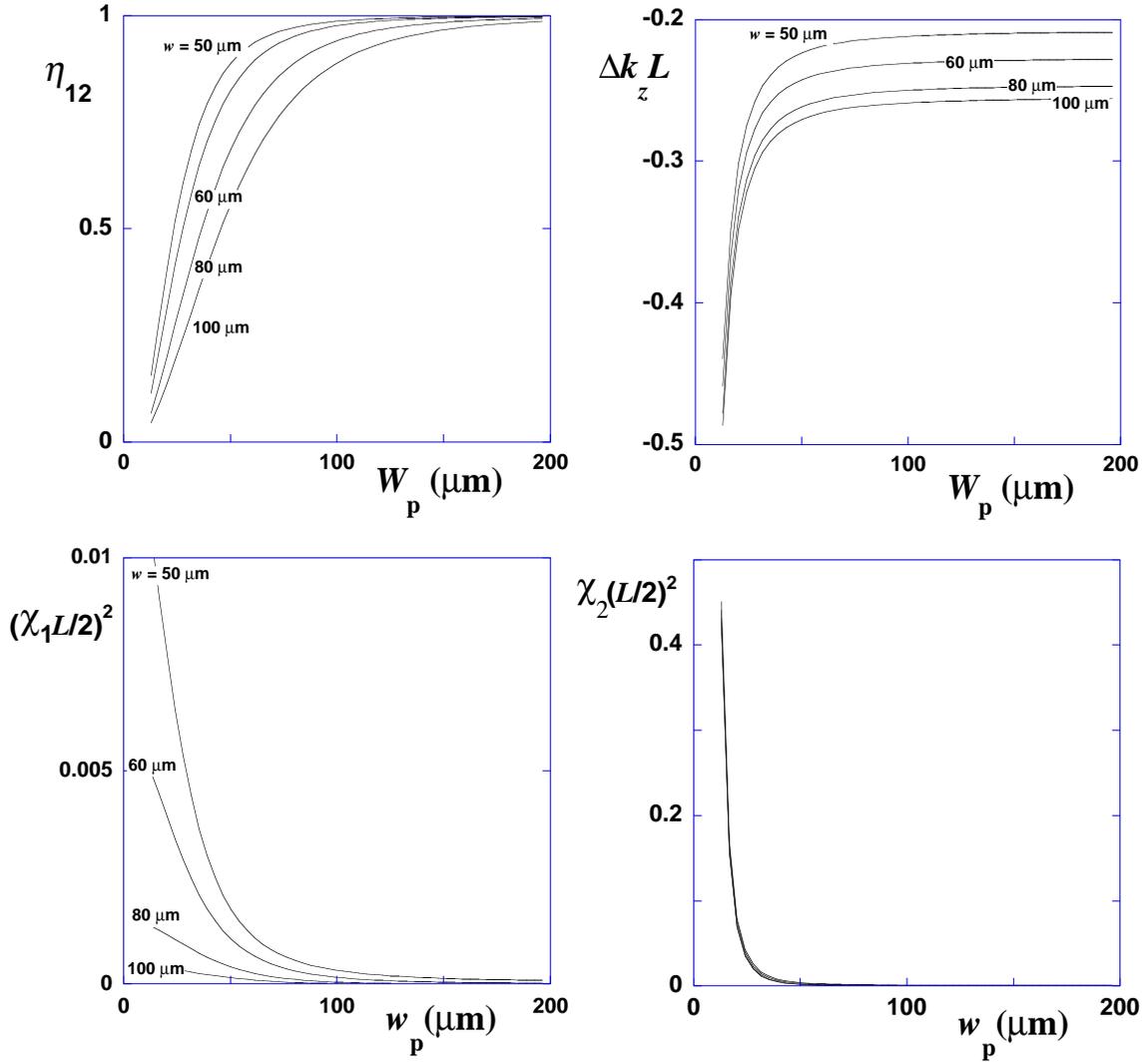}
\end{center}
\caption{ (a)Plot of the collection efficiency $\eta_{12}$
calculated versus the pump waist for various collection waists:
$\mathcal{W}=50~\mu\mathrm{m}$, 60~$\mu$m, 70 $\mu$m, 80 $\mu$m,
and 100 $\mu$m. The crystal length is $L=2~\mathrm{mm}$. (b) Plot
of the longitudinal mismatch $\Delta k_z L$ in the same condition
as case (a). (c) Plot of the correction factor $\chi_1^2 L^2/4$.
(d) Plot of $\chi_2 L^2/4$ --- the lines are so close that the
individual collection mode waists are not labelled.} \label{Figure
2}
\end{figure}

Figure~3 plots the collection efficiency versus the pump waist for
four different crystal lengths: $L= 0.1~\mathrm{mm}$, 2~mm, 3.5~mm
and 4.5~mm, at various mode waists (all are in the valid range of
the approximation). Note that for the case of $L= 0.1~\mathrm{mm}$
(which satisfies the thin crystal approximation) the optimum
collection efficiency can be obtained for all the collection modes
by picking a sufficiently large pump waist. However for the longer
crystals this is not the case.  For example, if the collection
mode is defined with $\mathcal{W}=20~\mu\mathrm{m}$ in the case of
$L= 4.5 ~ \mathrm{mm}$ then the maximum collection efficiency that
can be obtained is $\eta_{12} \approx 0.9$ regardless of how the
pump waist is varied.  However if we choose a collection mode with
$\mathcal{W}=50~\mu \mathrm{m}$ the collection efficiency is
$\eta_{12} \approx 1$ when $\mathcal{W}_\mathrm{p} > 100~\mu
\mathrm{m}$. (We limit the analysis here to crystal length of 4.5
mm because longer crystal push the validity of the approximations
made earlier.) From Fig.~3 we conclude that we can take advantage
of the higher signal levels from long crystals without sacrificing
coupling efficiency provided that the pump waist and collection
mode are scaled to appropriate values.
\begin{figure}[tbp]
\par
\begin{center}
\includegraphics[angle=0, width=15 cm, height=9 cm]{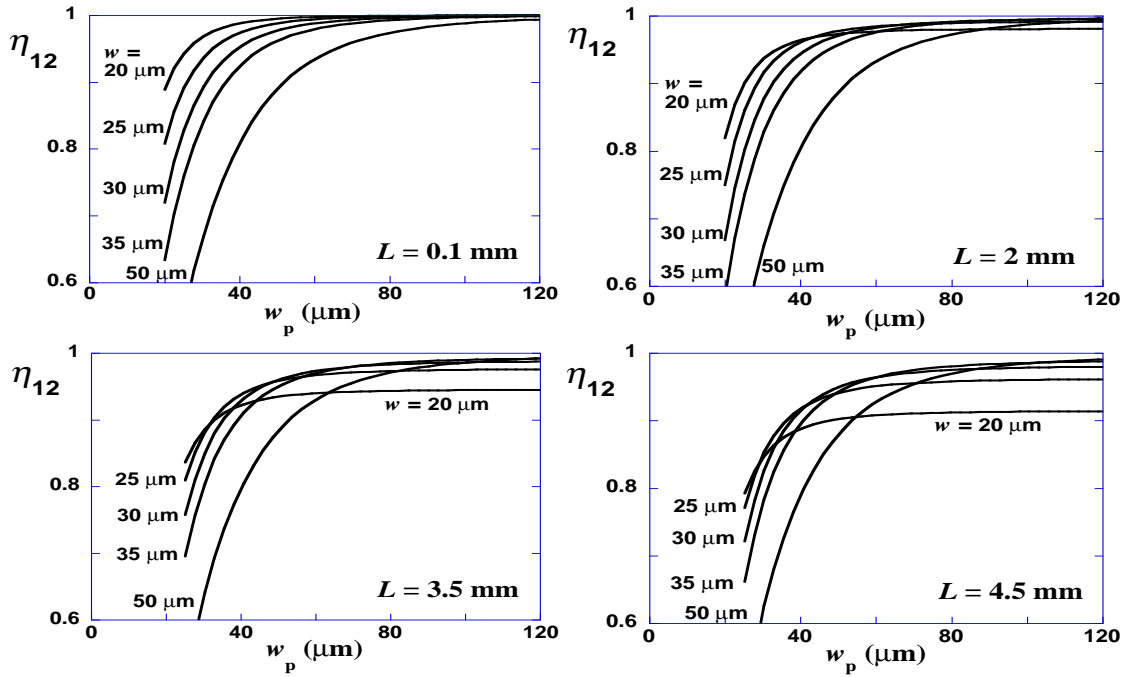}
\end{center}
\caption{Plot of the collection efficiency $\eta_{12}$ calculated
versus the pump waist  for various crystal length from left to
right $L=$0.1, $L$=2, $L$=3.5 and $L$=4.5 mm. Each graphics is
evaluated for collection mode waists $\mathcal{W}=20~\mu
\mathrm{m}$, 25~$\mu$m, 30~$\mu$m, 35 $\mu$m, 50 $\mu$m.}
\label{Figure 3}
\end{figure}

In Fig. 4a we plot the collection efficiency versus the crystal
length for different magnifications (M=15, 20, 25, 30, 40) while
maintaining a constant pump waist ($\mathcal{W}_\mathrm{p}=50~\mu
\mathrm{m}$) and fiber size ($\mathcal{W}_f=1.5 ~ \mu\mathrm{m}$).
Figure 4b is the same as Fig.~4a but in the non-perfect imaging
condition $\epsilon=0.4 \mathrm{m}^{-1}$, which reduces the
collection efficiency. Notice that as the crystal gets longer
larger collection modes give a better coupling efficiency. Even
though a longer crystal length reduces the collection efficiency
obtainable for a given pump size, these longer crystals may still
be desirable since they produce more overall signal.  In addition,
if the pump beam size is also be adjusted (as was done in the
Fig.~2) it is still possible to get good coupling efficiency with
a longer crystal.

\begin{figure}[tbp]
\par
\begin{center}
\includegraphics{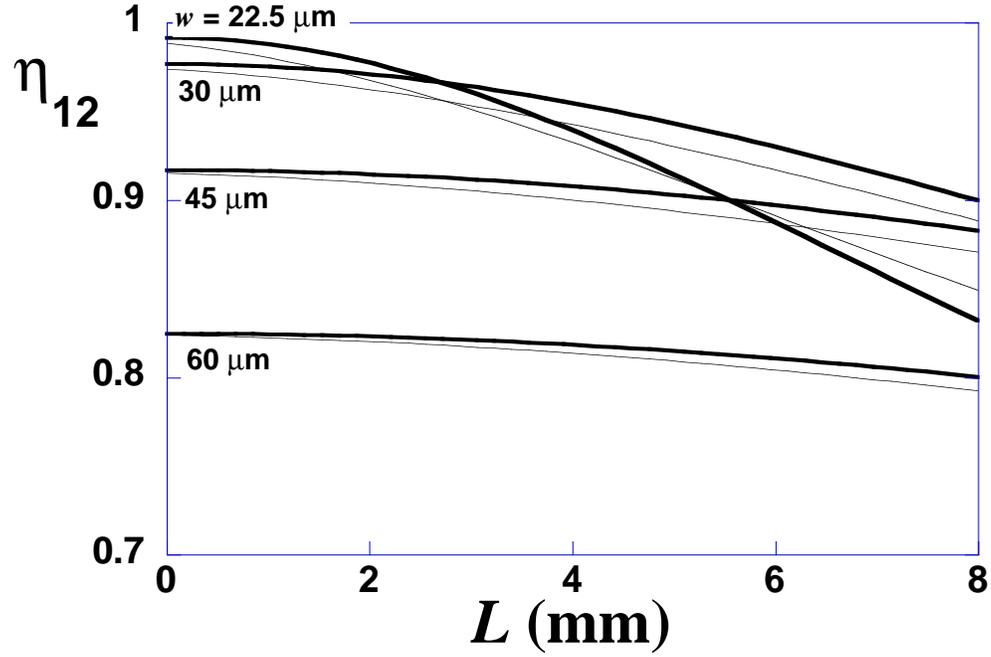}
\end{center}
\caption{(a)Plot of the collection efficiency $\eta_{12}$
calculated versus the crystal length at
$\mathcal{W}_\mathrm{p}$=50 $\mu$m for various collection mode
waists. Heavy lines are calculated for perfect imaging
($\epsilon=0$) and narrow lines are calculated for  the case of
$\epsilon=0.4 \mathrm{m}^{-1}$.} \label{Figure 3}
\end{figure}

To verify that Eq.~(\ref{eqn:efficiency}) is valid for the
parameters that we have chosen, we must consider the numerical
values of the corrections and verify that our approximation is in
its range of validity. This is shown in Fig.~5 where the
correction factors $\chi_1^2 L^2/4$ and $\chi_2 L^2/4$ are plotted
versus the crystal length for different magnifications (M=15, 20,
25, 30, 40), maintaining the pump waist at
$\mathcal{W}_\mathrm{p}=$50 $\mu$m and assuming the fiber core of
$\mathcal{W}_f$=$1.5 ~ \mu$m. It is clear that for the chosen
parameters for crystals longer than about 8 mm the corrections
become non-negligible.
\begin{figure}[tbp]
\par
\begin{center}
\includegraphics{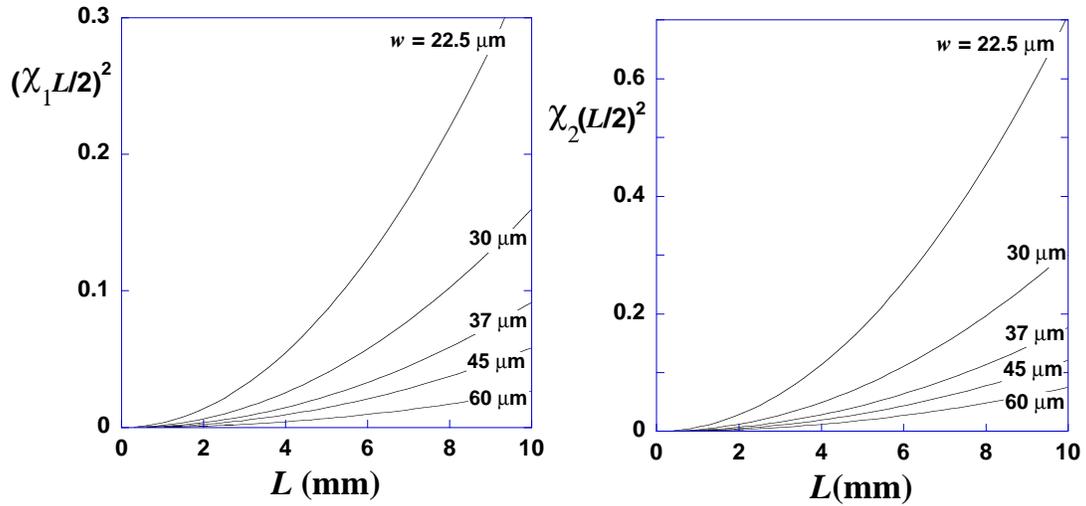}
\end{center}
\caption{Plot of the correction factors $\chi_1^2 L^2/4$ and
$\chi_2 L^2/4$ versus the crystal length for different collection
mode waists, maintaining the pump waist at
$\mathcal{W}_\mathrm{p}=$50 $\mu$m. } \label{Figure 3}
\end{figure}

\section{\label{sec:level1} Conclusions }
In the interest of optimizing the brightness of two-photon sources
for use in emerging applications of quantum information, we have
presented an analytical calculation of the collection efficiency
of PDC light into single mode fibers. Our calculation was
performed for the case of long pump pulse and type I PDC. These
experimental conditions are similar to those used in many SPOD
sources. We demonstrated that for crystals that are longer than
those required by the thin crystal approximation (but still within
the approximation made in the model) the maximum collection
efficiency can be obtained at larger pump waists. However, as
longer crystals are used the maximum achievable collection
efficiency decreases. The approximations made in obtaining the
expression for coupling efficiency limit the range of parameters
that can be studied using this method. It is particularly
important to pay close attention to the validity range of the
approximations to assure meaningful results, as this has not been
emphasized in some other works. Future work will move beyond the
simple two-photon efficiency analysis to include an overall signal
estimate that accounts for the effect of finite available pump on
the total signal received. Experimental tests will be conducted to
verify these analytical predictions.

This work was supported in part by DARPA/QUIST.


\begin{thebibliography}{10}

\bibitem{BEB84}
C.~Bennett and G.~Brassard, ``Quantum cryptography: public key
distribution and
  coin tossing,'' in {\em Proceedings of the IEEE International Conference on
  Computers, Systems and Signal Processing},  p.~175, 1984.

\bibitem{BEB85}
C.~Bennett and G.~Brassard, ``Quantum public key distribution
system,'' {\em
  IBM Technical Disclosure Bulletin} {\bf 28}, p.~3153, 1985.

\bibitem{BEB89}
C.~Bennett and G.~Brassard, ``The dawn of a new era for quantum
cryptography:
  The experimental prototype is working!,'' {\em SIGACT NEWS} {\bf 20}, p.~78,
  1989.

\bibitem{BBB91}
C.~H. Bennett, F.~Bessette, G.~Brassard, L.~Salvail, and
J.~Smolin,
  ``Experimental quantum cryptography,'' {\em Lecture Notes in Computer
  Science} {\bf 473}, pp.~253--265, 1991.

\bibitem{EKE91}
A.~Ekert, ``Quantum cryptography based on {B}ell's theorem,'' {\em
Phys. Rev.
  Lett.} {\bf 67}, pp.~661--663, 1991.

\bibitem{BEN92}
C.~Bennett, ``Quantum cryptography using any two nonorthogonal
states,'' {\em
  Phys. Rev. Lett.} {\bf 68}, pp.~3121--3124, 1992.

\bibitem{BBM92}
C.~H. Bennett, G.~Brassard, and N.~D. Mermin, ``Quantum
cryptography without
  {B}ell theorem,'' {\em Phys. Rev. Lett.} {\bf 68}(5), pp.~557--559, 1992.

\bibitem{ERT92}
A.~K. Ekert, J.~G. Rarity, P.~R. Tapster, and G.~M. Palma,
``Practical quantum
  cryptography based on 2-photon interferometry,'' {\em Phys. Rev. Lett.} {\bf
  69}(9), pp.~1293--1295, 1992.

\bibitem{TBZ00}
W.~Tittel, J.~Brendel, H.~Zbinden, and N.~Gisin, ``Quantum
cryptography using
  entangled photons in energy-time {B}ell states,'' {\em Phys. Rev. Lett.} {\bf
  84}(20), pp.~4737--4740, 2000.

\bibitem{KLM01}
E.~Knill, R.~Laflamme, and G.~J. Milburn, ``A scheme for efficient
quantum
  computation with linear optics,'' {\em Nature} {\bf 409}, pp.~46--52, 2001.

\bibitem{BLM00}
G.~Brassard, N.~Lutkenhaus, T.~Mor, and B.~C. Sanders,
``Limitations on
  practical quantum cryptography,'' {\em Phys. Rev. Lett.} {\bf 85}(6),
  pp.~1330--1333, 2000.

\bibitem{PJF02}
T.~B. Pittman, B.~C. Jacobs, and J.~D. Franson, ``Single photons
on
  pseudo-demand from stored parametric down-conversion,'' {\em Phys. Rev. A}
  {\bf 66}, pp.~042303 1--7, 2002.

\bibitem{MBC02}
A.~Migdall, D.~Branning, and S.~Castelletto, ``Tailoring
single-photon and
  multiphoton probabilities of a single-photon on-demand source,'' {\em Phys.
  Rev. A} {\bf 66}, pp.~053805 1--4, 2002.

\bibitem{MRP98a}
C.~H. Monken, P.~H.~S. Ribeiro, and S.~Padua, ``Optimizing the
photon pair
  collection effeciency: A step step towards a loophole-free {B}ell's
  inequalities experiment,'' {\em Phys. Rev. A} {\bf 57}, pp.~R2267--R2269,
  1998.

\bibitem{KOW01}
C.~Kurtsiefer, M.~Oberparleiter, and H.~Weinfurter,
``High-efficiency entangled
  photon pair collection in type-ii parametric fluorescence,'' {\em Phys. Rev.
  A} {\bf 6402}(2), p.~023802, 2001.

\bibitem{BGS03}
F.~A. Bovino, P.~Varisco, A.~M. Colla, G.~Castagnoli, G.~D.
Giuseppe, and A.~V.
  Sergienko, ``Effective fiber coupling of entangled photons for quantum
  communication,'' {\em ArXiv:quant-ph/0303126} , 2003.

\bibitem{RUB96}
M.~H. Rubin, ``Transverse correlation in optical spontaneous
parametric
  down-conversion,'' {\em Phys. Rev. A} {\bf 54}(6), pp.~5349--5360, 1996.

\end{thebibliography}

\end{document}